\documentclass{article}
\usepackage{spconf,amsmath,graphicx,hyperref,booktabs, amssymb, color, subcaption}

\title{
DNS: Data-driven Nonlinear Smoother for Complex Model-free Process}
\name{Fredrik Cumlin \quad
Anubhab Ghosh \quad Saikat Chatterjee}
\address{School of Electrical Engineering and Computer Science, KTH Royal Institute of Technology, Sweden
\\
\textit{fcumlin@kth.se, anubhabg@kth.se, sach@kth.se}
}
\begin{document}
\ninept
\maketitle
\begin{abstract}
We propose data-driven nonlinear smoother (DNS) to estimate a hidden state sequence of a complex dynamical process from a noisy, linear measurement sequence. The dynamical process is model-free, that is, we do not have any knowledge of the nonlinear dynamics of the complex process. There is no state-transition model (STM) of the process available. The proposed DNS uses a recurrent architecture that helps to provide a closed-form posterior of the hidden state sequence given the measurement sequence. DNS learns in an unsupervised manner, meaning the training dataset consists of only measurement data and no state data. We demonstrate DNS using simulations for smoothing of several stochastic dynamical processes, including a benchmark Lorenz system. Experimental results show that the DNS is significantly better than a deep Kalman smoother (DKS) and an iterative data-driven nonlinear state estimation (iDANSE) smoother.      
\end{abstract}
\section{Introduction}

Bayesian smoothing for a complex dynamical process is challenging, where the task is to compute the posterior of a hidden state sequence from a noisy measurement sequence. It is crucial for many applications such as denoising, object track estimation, path planning, robotics, and biomedical signal processing \cite{Särkkä_2013, smoothing_application, smoothing_application2}. The classical Rauch-Tung-Striebel (RTS) smoother is optimal for linear Gaussian systems \cite{rts}, and there exist particle smoothers for nonlinear systems based on ideas drawn from particle filters \cite{pf_convergence}. The RTS and particle smoothers require a priori knowledge of the underlying process to regularize the smoothing problem. They used a state-transition model (STM) that represents the state evolution of the dynamical process. Typically, an STM is derived using domain knowledge, such as the physics of the process. The use of first-order Markovian STMs is popular for various dynamical processes in various applications. 

In this article, we focus on smoothing for complex model-free processes. A complex process can be stochastic and highly nonlinear. We do not have good knowledge of the physics of the process, nor do we have an STM of the process. The process may have long-term memory, and the assumption of using a Markovian STM has limitations. 
The absence of an STM limits the use of RTS and particle smoothers.


For a model-free process, instead of STM-based smoothing methods (such as RTS and particle smoothers), we require data-driven methods based on machine learning for smoothing. Data-driven methods learn from training data. If we have access to labeled training data as pairwise measurement-and-state data, then supervised learning is an option. It is challenging to design methods based on unsupervised learning using unlabeled training data comprising measurement data and no access to state data. 

Recently, a hybrid smoother called RTSNet was proposed that combines STM-based model-driven RTS with data-driven machine learning, and learns in a supervised manner using labeled training data \cite{rtsnet}. 
Relevant unsupervised learning-based data-driven methods include dynamical variational autoencoders (DVAEs) \cite{dvae, dks} and differentiable particle filters (DPFs) \cite{dpf_overview, dpf_ot}. They typically use neural network-based Markovian STMs. A version of DVAE is called the deep Kalman smoother (DKS), which is particularly suited for smoothing, accounting for a suitable measurement system model.
There exists a data-driven nonlinear state estimation (DANSE) method which performs state estimation of a model-free process and learns in an unsupervised manner using measurement-only training data \cite{danse}. DANSE performs state estimation by maintaining a temporal causality similar to a filter. DANSE was extended for smoothing using an iterative mechanism, and the smoother is called iterative DANSE (iDANSE) \cite{iDANSE}. The iDANSE also maintains temporal causality and does not use the anti-causal part of a measurement sequence. For smoothing, it is important to use a long measurement sequence appropriately.

Our contribution in this article is to propose a data-driven nonlinear smoother (DNS) method for a model-free process in unsupervised learning settings using measurement-only training data. Like iDANSE, we use a linear measurement system model, and the DNS provides a closed-form posterior of the hidden state sequence. For the DNS, we develop a deep recurrent architecture (DRA) that can handle a long measurement sequence efficiently. The DRA comprises dense layers and gated recurrent units (GRUs) \cite{gru}, and has residual connections. GRUs are efficient recurrent neural networks (RNNs) that are unrolled and used in DNS. We evaluated the performance of DNS for the smoothing of three stochastic dynamical processes through simulations. The processes are: a stochastic Lorenz system, a stochastic Chen system, and a stochastic double spring-pendulum system. The first two are Markovian processes, and the third is a non-Markovian process. The DNS is compared with an extended RTS smoother that knows the STM of an underlying process, and data-driven DKS and iDANSE methods that do not have access to the STM of the underlying process. We show that DNS provides a significantly better performance than its competitors.

\section{Problem formulation and Background}

\begin{figure*}[htbp]
\centering
\begin{subfigure}[b]{0.48\textwidth}
    \centering
    \includegraphics[width=\textwidth]{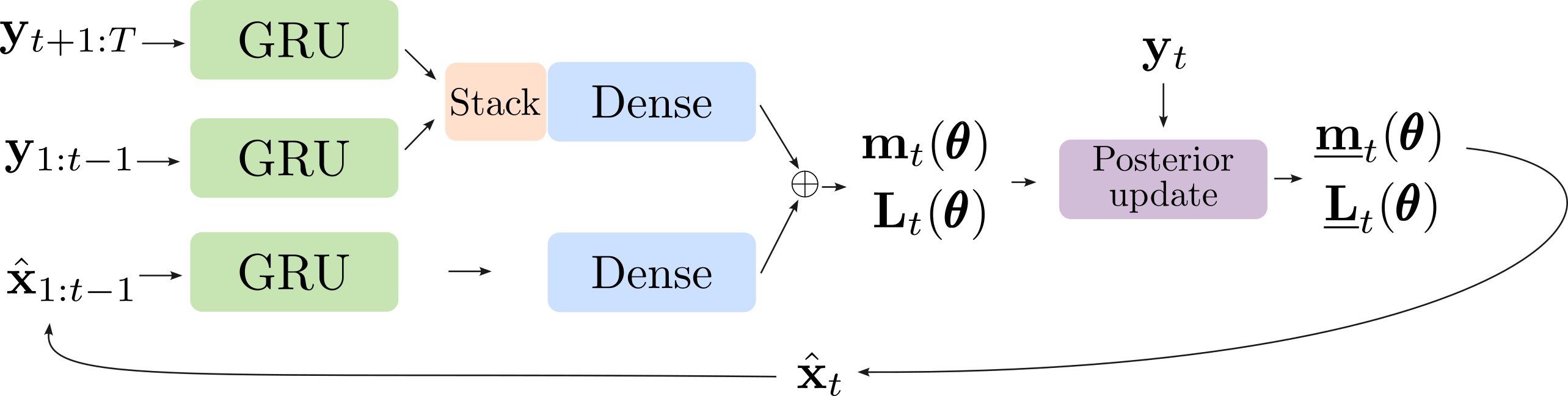}
    \caption{Deep recurrent architecture (DRA) of the proposed DNS}
    \label{fig:method_DNS}
\end{subfigure}
\hfill
\begin{subfigure}[b]{0.48\textwidth}
    \centering
    \includegraphics[width=\textwidth]{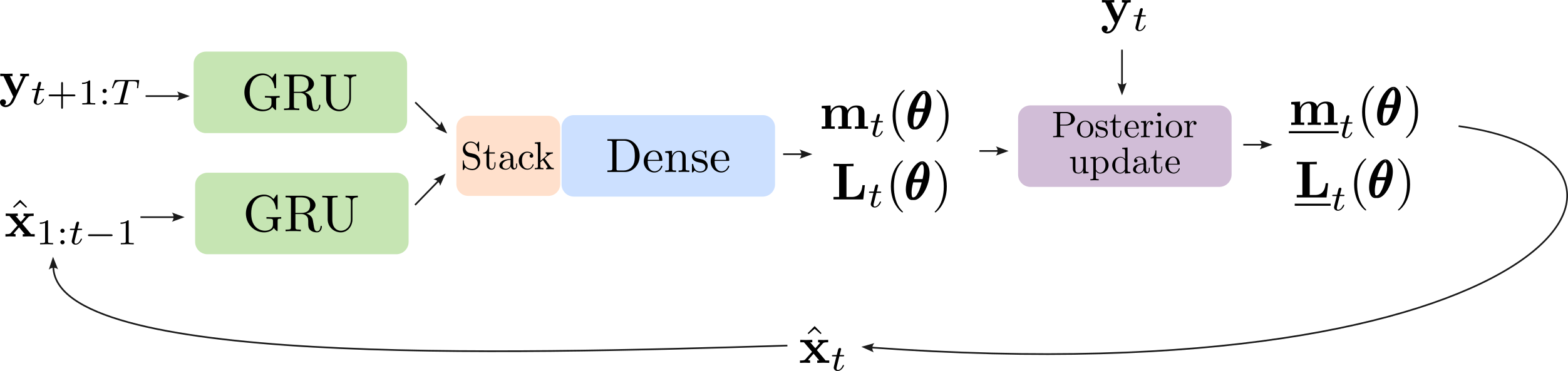}
    \caption{Lighter DRA of the simplified DNS}
    \label{fig:method_DNS_Simple}
\end{subfigure}
\caption{Deep recurrent architectures of the proposed data-driven nonlinear smoother (DNS)}
\end{figure*}

\subsection{Problem formulation}
Let $\mathbf{x}_t\in \mathbb{R}^{m}$ denotes the state vector of a model-free process $p(\{\mathbf{x}_t\})$ where $t\in\mathbb{N}_{+}$ is a discrete time-point index. The process generates state sequences; for example, a state sequence of length $T$, denoted as $\mathbf{x}_{1:T} \triangleq \mathbf{x}_1,\mathbf{x}_2,\ldots,\mathbf{x}_t, \ldots, \mathbf{x}_T \sim p(\{\mathbf{x}_t\})$. We consider a linear measurement model observing the state sequence:
\begin{eqnarray}
\label{eq:LinearMeasurementModel}
\mathbf{y}_t=\mathbf{H}\mathbf{x}_t+\mathbf{w}_t \in \mathbb{R}^{n},
\end{eqnarray}
where $\mathbf{H}$ is an ${n\times m}$ measurement matrix, $n \geq m$, and $\mathbf{w}_t\sim\mathcal{N}(\mathbf{w}_t; \mathbf{0},\mathbf{C}_w)$ is the additive Gaussian measurement noise. We assume that we know the measurement matrix $\mathbf{H}$ and the noise covariance matrix $\mathbf{C}_w$. For the state sequence $\mathbf{x}_{1:T}$, the measurement sequence is $\mathbf{y}_{1:T} \triangleq \left(\mathbf{y}_1,\mathbf{y}_2,\ldots,\mathbf{y}_t, \ldots, \mathbf{y}_T\right)$. 

The smoothing task is to find the posterior  $p\left(\mathbf{x}_{1:T}\vert\mathbf{y}_{1:T}\right)$. For the model-free process, we do not have an STM. Therefore, we are unable to use RTS and particle smoothers. Instead of an STM, we have access to a measurement-only unlabeled training dataset $\mathcal{D}$ to develop data-driven methods, as follows: 
\begin{eqnarray}
  \mathcal{D}=\left\{\mathbf{y}_{1:T^{(i)}}^{(i)}\right\}_{i=1}^N.  
\end{eqnarray}
The training dataset has $N$ independent measurement sequences. 
The $i$'th measurement sequence $\mathbf{y}_{1:T^{(i)}}^{(i)}$ has $T^{(i)}$-length; the measurement sequence corresponds to a state sequence $\mathbf{x}_{1:T^{(i)}}^{(i)}$ of $T^{(i)}$-length. Note that the training dataset has no access to state data, and we cannot develop a data-driven method based on supervised learning. We need to develop unsupervised learning-based methods.

\section{DNS: Data-driven Nonlinear Smoother}

In pursuit of the posterior $p\left(\mathbf{x}_{1:T}\vert\mathbf{y}_{1:T}\right)$, we develop the DNS where the following approximate decomposition is used:
\begin{equation}
\label{eq:smoothing_probability}
\begin{array}{rl}
p\left(\mathbf{x}_{1:T}\vert\mathbf{y}_{1:T}\right) \!\!\!\! &= p\left(\mathbf{x}_1\vert\mathbf{y}_{1:T}\right)\prod_{t=2}^T p\left(\mathbf{x}_{t}\vert \mathbf{x}_{1:t-1}, \mathbf{y}_{1:T}\right) \\
&\approx p\left(\mathbf{x}_1\vert\mathbf{y}_{1:T}\right)\prod_{t=2}^T p\left(\mathbf{x}_{t} \vert  \hat{\mathbf{x}}_{1:t-1}, \mathbf{y}_{1:T}\right).
\end{array}
\end{equation}
The approximation is in the second line of the above decomposition, where the true state sequence $\mathbf{x}_{1:t-1} \triangleq \mathbf{x}_1, \mathbf{x}_2, \ldots, \mathbf{x}_{t-1}$ is approximated by an estimated state sequence $\hat{\mathbf{x}}_{1:t-1} \triangleq \hat{\mathbf{x}}_1, \hat{\mathbf{x}}_2, \ldots, \hat{\mathbf{x}}_{t-1}$, where $\hat{\mathbf{x}}_{\tau} = \int_{\mathbf{x}_{\tau}} \mathbf{x}_{\tau}   p(\mathbf{x}_{\tau}|\hat{\mathbf{x}}_{1:\tau -1},\mathbf{y}_{1:T}) \, d\mathbf{x}_{\tau}$. The reason for the approximation is that the true state is hidden and not available. The DNS starts from $t=1$, computes $p(\mathbf{x}_1|\mathbf{y}_{1:T})$, and then proceeds to compute $p\left(\mathbf{x}_{t} \vert  \hat{\mathbf{x}}_{1:t-1}, \mathbf{y}_{1:T}\right)$ for $t=2, 3, \ldots, T$, sequentially.  

The use of \eqref{eq:smoothing_probability} in DNS overcomes a major shortcoming of the iDANSE smoother. iDANSE is based on DANSE, which is a state estimation method to find the posterior $p(\mathbf{x}_t|\mathbf{y}_{1:t})$. The DANSE is a filtering method that maintains causality like a Kalman filter. Extending DANSE for smoothing, the iDANSE also invokes causality as follows (see Eq. (3) in \cite{iDANSE}):
\begin{equation}
\label{eq:smoothing_iDANSE}
\begin{array}{rl}
p\left(\mathbf{x}_{1:T}\vert\mathbf{y}_{1:T}\right) &= p\left(\mathbf{x}_1\vert\mathbf{y}_{1:T}\right)\prod_{t=2}^T p\left(\mathbf{x}_{t}\vert \mathbf{x}_{1:t-1}, \mathbf{y}_{1:T}\right) \\
&\approx p\left(\mathbf{x}_1\vert\mathbf{y}_{1}\right)\prod_{t=2}^T p\left(\mathbf{x}_{t} \vert  \mathbf{x}_{1:t-1}, \mathbf{y}_{1:t}\right) \\
&\approx p\left(\mathbf{x}_1\vert\mathbf{y}_{1}\right)\prod_{t=2}^T p\left(\mathbf{x}_{t} \vert  \hat{\mathbf{x}}_{1:t-1}, \mathbf{y}_{1:t}\right).
\end{array}
\end{equation}
In the above decomposition for iDANSE, we maintain causality and explicitly use the approximation $p\left(\mathbf{x}_{t}\vert \mathbf{x}_{1:t-1}, \mathbf{y}_{1:T}\right) \approx p\left(\mathbf{x}_{t}\vert \mathbf{x}_{1:t-1}, \mathbf{y}_{1:t}\right)$. Therefore, iDANSE does not use the anti-causal part $\mathbf{y}_{t+1:T}$. The absence of the anti-causal part results in a loss of smoothing performance. The proposed DNS explicitly uses the anti-causal part. In the next subsection, we describe DNS and its unsupervised learning algorithm. 

\subsection{DNS and its unsupervised learning}

\begin{table*}[htbp]
\centering
\renewcommand{\arraystretch}{0.8}
\caption{Performance of various smoothers}
\footnotesize
\setlength{\tabcolsep}{1pt}
\begin{tabular}{l|ccc|ccc|ccc}
\toprule
\textbf{Process} $\rightarrow$ & \multicolumn{3}{c|}{\textbf{Stochastic Lorenz System}} & \multicolumn{3}{c|}{\textbf{Stochastic Chen System}} & \multicolumn{3}{c}{\textbf{Stochastic Double-Spring Pendulum}} \\ 

 & \multicolumn{3}{c|}{SMNR (dB)} & \multicolumn{3}{c|}{SMNR (dB)} & \multicolumn{3}{c}{SMNR (dB)} \\
\textbf{Model} $\downarrow$ & -10 & 0 & 10 & -10 & 0 & 10 & -10 & 0 & 10 \\
\midrule
\multicolumn{10}{c}{NMSE (dB)} \\
\midrule
\textit{ERTSS} & $\mathit{-2.56{\scriptstyle\pm 0.03}}$ & $\mathit{-12.75{\scriptstyle\pm 0.02}}$ & $\mathit{-23.47{\scriptstyle\pm 0.02}}$ & $\mathit{-4.72{\scriptstyle\pm 0.07}}$ & $\mathit{-13.13{\scriptstyle\pm 0.05}}$ & $\mathit{-21.52{\scriptstyle\pm 0.02}}$ & NA & NA & NA \\[0.5ex]
DANSE & $-6.07{\scriptstyle\pm 0.01}$ & $-13.93{\scriptstyle\pm 0.01}$ & $-21.24{\scriptstyle\pm 0.01}$ & $-3.07{\scriptstyle\pm 0.01}$ & $-9.83{\scriptstyle\pm 0.01}$ & $-18.66{\scriptstyle\pm 0.01}$ & $-5.85{\scriptstyle\pm 0.01}$ & $-10.19{\scriptstyle\pm 0.01}$ & $-16.53{\scriptstyle\pm 0.01}$ \\[0.5ex]
DKS & $-1.00{\scriptstyle\pm 0.17}$ & $-5.96{\scriptstyle\pm 0.22}$ & $-14.62{\scriptstyle\pm 0.14}$ & $-0.22{\scriptstyle\pm 0.57}$ & $-3.29{\scriptstyle\pm 0.14}$ & $-11.37{\scriptstyle\pm 0.39}$ & $-0.86{\scriptstyle\pm 0.06}$ & $-2.89{\scriptstyle\pm 0.09}$ & $-8.30{\scriptstyle\pm 0.07}$ \\[0.5ex]
iDANSE & $-5.94{\scriptstyle\pm 0.20}$ & $-13.93{\scriptstyle\pm 0.01}$ & $-21.91{\scriptstyle\pm 0.01}$ & $\mathbf{-3.85}{\scriptstyle\pm 0.77}$ & $-11.20{\scriptstyle\pm 0.22}$ & $\mathbf{-18.64}{\scriptstyle\pm 0.03}$ & $\mathbf{-5.99}{\scriptstyle\pm 0.06}$ & $-10.20{\scriptstyle\pm 0.06}$ & $-16.95{\scriptstyle\pm 0.06}$ \\[0.5ex]
DNS-S & $-2.49{\scriptstyle\pm 0.99}$ & $-14.37{\scriptstyle\pm 0.52}$ & $-22.24{\scriptstyle\pm 0.81}$ & $-1.92{\scriptstyle\pm 1.77}$ & $-10.38{\scriptstyle\pm 0.99}$ & $-17.71{\scriptstyle\pm 1.13}$ & $-3.73{\scriptstyle\pm 0.90}$ & $-10.18{\scriptstyle\pm 0.15}$ & $-17.17{\scriptstyle\pm 0.42}$ \\[0.5ex]
DNS \, & $\mathbf{-6.26}{\scriptstyle\pm 0.69}$ & $\mathbf{-16.84}{\scriptstyle\pm 0.31}$ & $\mathbf{-22.74}{\scriptstyle\pm 0.53}$ & $-3.42{\scriptstyle\pm 1.03}$ & $\mathbf{-12.54}{\scriptstyle\pm 0.65}$ & $-18.20{\scriptstyle\pm 1.06}$ & $-5.23{\scriptstyle\pm 1.25}$ & $\mathbf{-10.53}{\scriptstyle\pm 0.1}$ & $\mathbf{-17.18}{\scriptstyle\pm 0.16}$ \\
\midrule
\multicolumn{10}{c}{ALP} \\
\midrule
DKS & $-3.2\times 10^5$ & $-2.5\times 10^5$ & $-1.8\times 10^5$ & $-3.2\times 10^4$ & $-2.6\times 10^4$ & $-1.9\times 10^4$ & $-1.9\times 10^4$ & $-1.1\times 10^4$ & $-2.3\times 10^3$ \\[0.5ex]
iDANSE & $-2.5\times 10^{9}$ & $-43.01{\scriptstyle\pm 22.04}$ & $-5.17{\scriptstyle\pm 0.01}$ & $-1.1\times 10^{26}$ & $-97.35{\scriptstyle\pm 26.22}$ & $\mathbf{-6.88}{\scriptstyle\pm 0.24}$ & $-173.9{\scriptstyle\pm 308.6}$ & $-119.4{\scriptstyle\pm 253.9}$ & $-0.93{\scriptstyle\pm 8.67}$ \\[0.5ex]
DNS-S & $-1.2\times10^3$ & $-13.19{\scriptstyle\pm 8.68}$ & $-5.43{\scriptstyle\pm 0.57}$ & $-1.1\times10^{5}$ & $-53.47{\scriptstyle\pm 48.4}$ & $-10.88{\scriptstyle\pm 2.88}$ & $-10.15{\scriptstyle\pm 5.64}$ & $-1.92{\scriptstyle\pm 0.35}$ & $\mathbf{2.02}{\scriptstyle\pm 0.21}$ \\[0.5ex]
DNS \, & $\mathbf{-26.79}{\scriptstyle\pm 12.99}$ & $\mathbf{-7.28}{\scriptstyle\pm 0.44}$ & $\mathbf{-4.82}{\scriptstyle\pm 0.32}$ & $\mathbf{-88.95}{\scriptstyle\pm 90.31}$ & $\mathbf{-10.15}{\scriptstyle\pm 0.81}$ & $-7.07{\scriptstyle\pm 0.52}$ & $\mathbf{-5.97}{\scriptstyle\pm 5.07}$ & $\mathbf{-1.21}{\scriptstyle\pm 0.15}$ & $1.99{\scriptstyle\pm 0.08}$ \\
\bottomrule
\end{tabular}
\label{tab:model_performance}
\end{table*}

The development of DNS has two phases: an inference phase and a training phase. We first discuss the inference phase where we deal with providing the posterior following \eqref{eq:smoothing_probability} given the measurement sequence $\mathbf{y}_{1:T}$ and the linear measurement system \eqref{eq:LinearMeasurementModel}. The training phase will be explained later.

In the inference phase, we use a deep recurrent architecture (DRA) that we design to compute $\forall t, p\left(\mathbf{x}_{t} \vert \hat{\mathbf{x}}_{1:t-1}, \mathbf{y}_{1:T}\right) \triangleq p\left(\mathbf{x}_{t} \vert \hat{\mathbf{x}}_{1:t-1}, \mathbf{y}_{1:t-1}, \mathbf{y}_t, \mathbf{y}_{t+1:T}\right)$. Then we use \eqref{eq:smoothing_probability} to compute the full posterior $p\left(\mathbf{x}_{1:T}\vert\mathbf{y}_{1:T}\right)$. For a time-point $t$, the input to the DRA has three parts: the past measurement sequence $\mathbf{y}_{1:t-1}$, the anti-causal measurement sequence $\mathbf{y}_{t+1:T}$, and the past estimated state sequence $\hat{\mathbf{x}}_{1:t-1}$. Using the three parts as input, the DRA provides parameters of a Gaussian prior, as follows: 
\begin{eqnarray}
\label{eq:GaussianPrior_DNS}
\begin{array}{rl}
    p \!\left(\mathbf{x}_t\vert \hat{\mathbf{x}}_{1:t-1}, \! \mathbf{y}_{1:t-1},\!\mathbf{y}_{t+1:T}\right) \!\!\!\!\!\! & \triangleq \! p\!\left(\mathbf{x}_t\vert \hat{\mathbf{x}}_{1:t-1},\!\mathbf{y}_{1:t-1}, \!\mathbf{y}_{t+1:T};\pmb{\theta}\right) \\
     & =  \mathcal{N}\left(\mathbf{x}_t; \mathbf{m}_t\!\left(\pmb{\theta}\right), \mathbf{L}_t\!\left(\pmb{\theta}\right)\right).
\end{array}
\end{eqnarray}
The parameters of the Gaussian prior are its mean vector $\mathbf{m}_t\!\left(\pmb{\theta}\right)$ and covariance matrix $\mathbf{L}_t\!\left(\pmb{\theta}\right)$. Let the set of DRA parameters (for example, the weight matrices) be denoted by $\pmb{\theta}$. Then the output of the DRA is a function of $\pmb{\theta}$. That means we write:
\begin{eqnarray}\label{eq:prior_mapping}
    \mathrm{DRA}(\hat{\mathbf{x}}_{1:t-1},\mathbf{y}_{1:t-1}, \mathbf{y}_{t+1:T};\pmb{\theta}) \rightarrow \left\{ \mathbf{m}_t\!\left(\pmb{\theta}\right), \mathbf{L}_t\!\left(\pmb{\theta}\right) \right\}.
\end{eqnarray}

Using the Gaussian prior \eqref{eq:GaussianPrior_DNS} and the current measurement $\mathbf{y}_t$ that follows \eqref{eq:LinearMeasurementModel}, we can compute the posterior $p\left(\mathbf{x}_{t} \vert \hat{\mathbf{x}}_{1:t-1}, \mathbf{y}_{1:T}\right)$ in closed-form Gaussian applying `completing the square' approach \cite[Ch. 2]{bishop2006pattern} and Woodbury matrix identity, as follows:
\begin{eqnarray}
\label{eq:GaussianPosterior_DNS}
\begin{array}{rl}\label{eq:DNS_prior_to_posterior}
    p \left(\mathbf{x}_t\vert \hat{\mathbf{x}}_{1:t-1},  \mathbf{y}_{1:T}\right) \!\!\!\!\! & \triangleq p\left(\mathbf{x}_t\vert \hat{\mathbf{x}}_{1:t-1},\mathbf{y}_{1:T};\pmb{\theta}\right) \\
     & =  \mathcal{N}\left(\mathbf{x}_t; \underline{\mathbf{m}}_t \!\left(\pmb{\theta}\right), \underline{\mathbf{L}}_t \!\left(\pmb{\theta}\right)\right),
\end{array}
\end{eqnarray}
where $\underline{\mathbf{m}}_t \!\left(\pmb{\theta}\right)  =   \mathbf{m}_t \!\left(\pmb{\theta}\right) + \mathbf{K}_t \left(\mathbf{y}_t -\mathbf{H} \mathbf{m}_t\!\left(\pmb{\theta}\right) \right)$, $\underline{\mathbf{L}}_t \!\left(\pmb{\theta}\right) = \mathbf{L}_t\!\left(\pmb{\theta}\right) - \mathbf{K}_t \mathbf{R}_t \mathbf{K}_t^{\top}$, 
$\mathbf{K}_t  =\mathbf{L}_t\!\left(\pmb{\theta}\right) \mathbf{H}^\top \mathbf{R}_t^{-1}$ and $\mathbf{R}_t  =\mathbf{H} \mathbf{L}_t\!\left(\pmb{\theta}\right) \mathbf{H}^\top + \mathbf{C}_w$. A point estimate of $\mathbf{x}_t$, denoted $by$ $\hat{\mathbf{x}}_t$ can be chosen as $\hat{\mathbf{x}}_t = \underline{\mathbf{m}}_t \!\left(\pmb{\theta}\right)$. This procedure of providing a Gaussian prior followed by computing a Gaussian posterior is performed for $t=1,2,\ldots,T$ sequentially, starting from $t=1$.     

\subsubsection{DRA Architecture and Unsupervised Learning}

The DRA that we develop has three GRUs as RNNs and two dense layers to combine relevant information. The developed architecture is shown in Fig.~\ref{fig:method_DNS}. The future measurement sequence, past measurement sequence, and the estimated states are processed by GRUs, respectively. The outputs of three GRUs are stacked and processed by dense layers. The output of the GRU that uses the estimated state sequence $\hat{\mathbf{x}}_{1:t-1}$ as input is also processed by dense layers. These outputs are then added to predict the prior parameters as given in Eq~\ref{eq:prior_mapping}. The later incorporation of the information of the estimated states is called a skip connection. We will motivate the skip connection in an ablation study in the experimental section. Given the current measurement $\mathbf{y}_t$, the prior parameters are mapped to the parameters of the posterior as described in Eq.~\ref{eq:DNS_prior_to_posterior}.

Next, we discuss the unsupervised learning method that we developed to train the DRA of the DNS using the training dataset $\mathcal{D}$. Let us consider $t$'th time-point of measurement sequence $\mathbf{y}_{1:T}$. Given $\hat{\mathbf{x}}_{1:t-1}$, $\mathbf{y}_{1:t-1}$, and $\mathbf{y}_{t+1:T}$, the logarithmic likelihood of measurement $\mathbf{y}_t$ can be expressed using the following closed-form Gaussian relation (applying `completing the square' approach \cite[Ch. 2]{bishop2006pattern}):
\begin{equation}
\label{eq:trajectory_log_lilihood_method1}
\begin{array}{l}
    \log p( \mathbf{y}_t \vert \hat{\mathbf{x}}_{1:t-1}, \mathbf{y}_{1:t-1}, \mathbf{y}_{t+1:T};\pmb{\theta}) \\
    = \log \int_{\mathbf{x}_t} p\left(\mathbf{y}_t\vert \mathbf{x}_{t}\right)p\left(\mathbf{x}_t\vert \hat{\mathbf{x}}_{1:t-1}, \mathbf{y}_{1:t-1},\mathbf{y}_{t+1:T}; \pmb{\theta} \right)d\mathbf{x}_t \\
    = \log \int_{\mathbf{x}_t}  \mathcal{N} \left( \mathbf{y}_t; \mathbf{H}\mathbf{x}_t, \mathbf{C}_w  \right) \mathcal{N}\left(\mathbf{x}_t; \mathbf{m}_t\!\left(\pmb{\theta}\right), \mathbf{L}_t\!\left(\pmb{\theta}\right)\right) d\mathbf{x}_t \\
    = 
     \log \mathcal{N}\left(\mathbf{y}_t; \mathbf{H} \mathbf{m}_t\!\left(\pmb{\theta}\right), \mathbf{C}_w + \mathbf{H} \mathbf{L}_t\!\left(\pmb{\theta}\right) \mathbf{H}^\top\right)\\
    = -\frac{n}{2}\log 2\pi - \frac{1}{2}\log\det \left( \mathbf{C}_w + \mathbf{H} \mathbf{L}_t\!\left(\pmb{\theta}\right) \mathbf{H}^\top \right) \\
    \hspace{0.4cm} - \frac{1}{2} \| \mathbf{y}_t - \mathbf{H}\mathbf{m}_t\!\left(\pmb{\theta}\right)\|^2_{\left( \mathbf{C}_w + \mathbf{H} \mathbf{L}_t\!\left(\pmb{\theta}\right) \mathbf{H}^\top \right)^{-1}},
\end{array}
\end{equation}
where we use the notation $\| \mathbf{a} - \mathbf{b} \|^2_{\mathbf{C}^{-1}} = \left( \mathbf{a} - \mathbf{b} \right)^{\top} \mathbf{C}^{-1} \left( \mathbf{a} - \mathbf{b} \right)$ for two vectors $\mathbf{a}$ and $\mathbf{b}$, and covariance $\mathbf{C}$. In the second line of \eqref{eq:trajectory_log_lilihood_method1}, we use the relation $p\left(\mathbf{y}_t\vert \mathbf{x}_t, \hat{\mathbf{x}}_{1:t-1}, \mathbf{y}_{1:t-1},\mathbf{y}_{t+1:T}; \pmb{\theta} \right) = p(\mathbf{y}_t | \mathbf{x}_t)$. 
Then, using \eqref{eq:trajectory_log_lilihood_method1} and the training dataset $\mathcal{D}$, we form the following logarithmic likelihood:  
\begin{equation}
\label{eq:loss_method2}
    \mathcal{L}(\mathcal{D}; \pmb{\theta})=\sum_{i=1}^N\sum_{t=1}^{T^{(i)}} \log p \left( \mathbf{y}_t\vert \hat{\mathbf{x}}_{1:t-1}, \mathbf{y}_{1:t-1}, \mathbf{y}_{t+1:T}; \pmb{\theta}\right).
\end{equation}
Finally, the maximum likelihood principle-based optimization is:
\begin{eqnarray}
    \arg\max_{\pmb{\pmb{\theta}}} \mathcal{L}(\mathcal{D}; \pmb{\theta}).
\end{eqnarray}

Note that, due to the requirement to use the estimated state sequence $\hat{\mathbf{x}}_{1:t-1}$ to compute the parameters of the Gaussian prior and use the parameters $\left\{ \mathbf{m}_t\!\left(\pmb{\theta}\right), \mathbf{L}_t\!\left(\pmb{\theta}\right) \right\}$ to compute \eqref{eq:trajectory_log_lilihood_method1}, the inference phase of the training sequences is embedded in the learning phase. 

\subsubsection{On a simple approach}
\label{subsubsec:DNS-S}

Typically, for a dynamical process, given the past state sequence $\mathbf{x}_{1:t-1}$, the current state $\mathbf{x}_t$ is conditionally independent of the past measurement sequence $\mathbf{y}_{1:t-1}$; that is, $\mathbf{x}_t \perp \mathbf{y}_{1:t-1}\mid \mathbf{x}_{1:t-1}$. Therefore, we could simplify \eqref{eq:smoothing_probability} as follows:
\begin{eqnarray}
\label{eq:smoothing_probability_simple}
\begin{array}{rl}
p\left(\mathbf{x}_{1:T}\vert\mathbf{y}_{1:T}\right) \!\!\!\!\! &= p\left(\mathbf{x}_1\vert\mathbf{y}_{1:T}\right)\prod_{t=2}^T p\left(\mathbf{x}_{t}\vert \mathbf{x}_{1:t-1}, \mathbf{y}_{1:T}\right) \\
&= p\left(\mathbf{x}_1\vert\mathbf{y}_{1:T}\right)\prod_{t=2}^T p\left(\mathbf{x}_{t}\vert \mathbf{x}_{1:t-1},  \mathbf{y}_{t:T}\right) \\
&\approx p\left(\mathbf{x}_1\vert\mathbf{y}_{1:T}\right)\prod_{t=2}^T p\left(\mathbf{x}_{t} \vert  \hat{\mathbf{x}}_{1:t-1}, \mathbf{y}_{t:T}\right),
\end{array}
\end{eqnarray}
where we discard the past sequence part $\mathbf{y}_{1:t-1}$. This simplification could help to design a lighter DRA that no longer uses $\mathbf{y}_{1:t-1}$ as input. An architecture of the lighter DRA is shown Fig.~\ref{fig:method_DNS_Simple}. Such a simplification requires the assumption that  $p\left(\mathbf{x}_t\vert \hat{\mathbf{x}}_{1:t-1}, \mathbf{y}_{1:T}\right)\approx p\left(\mathbf{x}_t\vert \hat{\mathbf{x}}_{1:t-1}, \mathbf{y}_{t:T}\right)$. In our experimental study, we later show that such an assumption results in a loss of smoothing performance when the power of the measurement noise $\mathbf{w}_t$ is high. In that case, the (full) DNS provides a significantly better performance than the simplified DNS based on the lighter DRA.

\section{Experiments and results}

Using simulations, we evaluate DNS based on synthetic data generated from three stochastic dynamical systems - a three-dimensional stochastic Lorenz system with $\mathbf{x}_t \in \mathbb{R}^3$, a three-dimensional Chen system with $\mathbf{x}_t \in \mathbb{R}^3$, and two-dimensional positions of two balls of a stochastic double-spring pendulum (SDSP) with $\mathbf{x}_t \in \mathbb{R}^4$. For SDSP, two balls represent the tip points of the two pendulums. The Lorenz and Chen systems are Markovian, and the four-dimensional position of the two balls of SDSP is non-Markovian.

The STMs that we simulate for the stochastic Lorenz and Chen systems are shown in \cite{danse}; see Eq. (18-19) of \cite{danse} for the stochastic Lorenz system, and Eq. (21-22) of \cite{danse} for the stochastic Chen system. We used the same parameter settings mentioned in \cite{danse} for the stochastic Lorenz and Chen systems. The SDSP is a complex system; it has eight internal state variables to characterize the dynamics of the SDSP. The eight internal state variables are: two angular positions, their changes over time
as angular velocities, lengths of the two springs, and changes in the lengths over time. We simulate a Markovian model for the eight internal state variables using the Runge-Kutta method. The dynamics of the position of the two balls together form a non-Markovian four-dimensional state vector for SDSP that we experiment with.

\begin{figure}[t]
\centering
\includegraphics[width=0.48\textwidth]{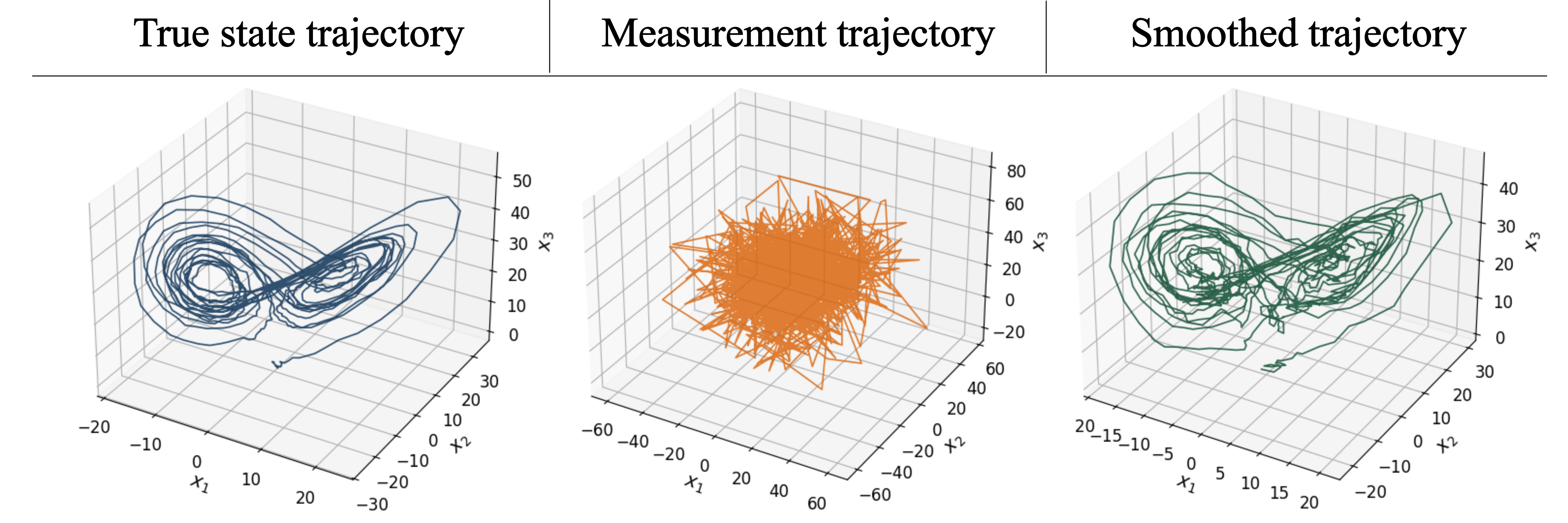}
\caption{Lorenz-63 demonstration with DNS at $0$ dB}
\label{fig:DNS_demonstration_for_Lorenz}
\end{figure}

We compare DNS with an extended Rauch-Tung-Streibler smoother (ERTSS) \cite{rts, jazwinski1970stochastic}, DANSE \cite{danse}, DKS (with the known measurement system \eqref{eq:LinearMeasurementModel}) \cite{dks}, and iDANSE \cite{iDANSE}. ERTSS uses STM of the underlying dynamical process. Therefore, we did not implement ERTSS for SDSP system due to the absence of a suitable STM. The DANSE, DKS, and iDANSE are data-driven methods. We use $\mathbf{H}=\mathbf{I}$ for all our experiments, that means $n=m$, and isotropic Gaussian measurement noise $\mathbf{w}_t\sim\mathcal{N}(\mathbf{w}_t; \mathbf{0},\sigma_w^2 \mathbf{I})$. For each dynamical system, we generate sequences using simulations - state and measurement sequences. The training phase uses short measurement sequences of length 100, and the inference phase (testing phase) uses long sequences of length 1000. The use of long sequences in the inference phase helps to observe how the data-driven smoothers generalize. For training, we use $N=1000$ sequences, and for testing, we use $N_{test}=100$ sequences. We tested the methods in three different signal-to-measurement noise ratios (SMNRs): $-10$, $0$, and $10$ dB. By choosing a value of \( \sigma^2_w \), for $N$ measurement sequences, SMNR in dB is:
$\mathrm{SMNR} {=}  \frac{1}{N}\!\sum_{i=1}^{N} 10\log_{10}\sum_{t=1}^{T}\frac{\mathbb{E}\!\left[\!\left|\!\left|\mathbf{h}(\mathbf{x}_t^{(i)})\!-\mathbb{E}\left[{\mathbf{h}(\mathbf{x}_t^{(i)})}\right]\!\right|\!\right|^2_2\right]}{\text{tr}(\mathbf{C}_w)}$.
A high $\sigma^2_w$ leads to a low SMNR. For all experimental settings, we train $10$ realizations of a method and report the mean ± standard deviation results of the 10 realizations.

In our experiments, the DNS processes input signals through causal 1D convolutions ($16$ kernels of length $3$), GRU layers ($1$ layer, $30$ hidden units), and $3$-layer fully connected dense networks. During training, we used mini-batch gradient descent (batch size $64$) with Adam optimizer (lr$=10^{-3}$, reduced by $0.9$ every $33$ epochs) for $200$ epochs.\footnote{Code can be found at \url{https://github.com/fcumlin/DataDrivenNonLinearSmoother}.} We use two evaluation measures for testing:  normalized mean square error (NMSE) in dB and average log posterior probability (ALP). They are given by 
$\text{NMSE} = \frac{1}{N_{test}} \sum_{i=1}^{N_{test}} 10 \log_{10} \frac{\sum_{t=1}^{1000} \|\mathbf{x}_t^{(i)} - \hat{\mathbf{x}}_t^{(i)} \|_2^2}{\sum_{t=1}^{1000} \|\mathbf{x}_t^{(i)}\|_2^2}$ and $\text{ALP} = \frac{1}{N_{test}} \sum_{i=1}^{N_{test}} \frac{1}{1000} \sum_{t=1}^{1000} \log p(\mathbf{x}_t^{(i)} | \hat{\mathbf{x}}_{1:t-1}^{(i)},\mathbf{y}_{1:T}^{(i)})$, as used in \cite{danse} and \cite{alp_loss}, respectively. The NMSE measures the performance of the state estimation as a point estimate, whereas the ALP measures the performance of the posterior estimation. It is desired to have a low NMSE and a high ALP.

\subsection{Results}

\begin{figure}[t]
\centering
\includegraphics[width=0.47\textwidth]{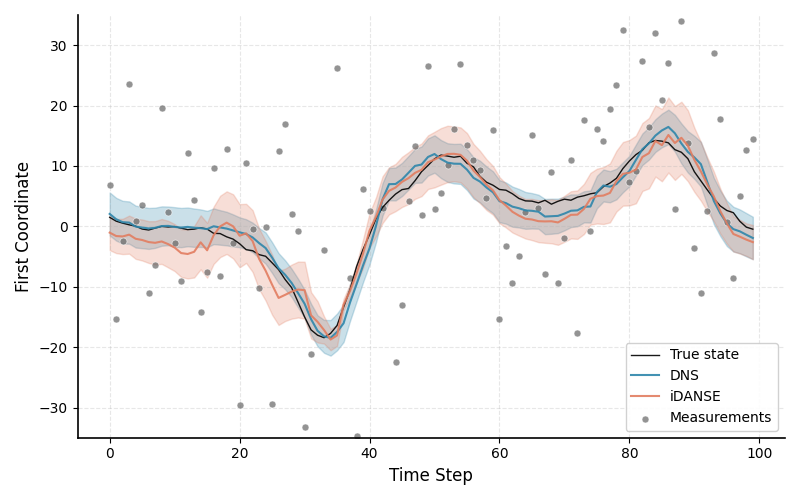}
\caption{Visualization of state estimation and uncertainty quantification for the stochastic Lorenz system at 0 dB SMNR. The trajectories for the first coordinate of the three-dimensional state are shown.}
\label{fig:uncertainty_comparison}
\end{figure}

The NMSE and ALP performance results are presented in Table~\ref{tab:model_performance}. In the table, DNS-S represents the simplified DNS discussed in Section \ref{subsubsec:DNS-S}. We first compare DNS with DNS-S. At high SMNR levels, they perform similarly, suggesting that the conditional independence assumption holds when the measurements are less noisy. As the SMNR decreases, DNS outperforms DNS-S for all three dynamical systems, suggesting that the conditional independence using estimated state sequence breaks down with noisy measurements. This breakdown is more evident for ALP performance, and the posterior estimation of DNS-S is poorer than DNS for low SMNRs. 

Next, we compare DNS with the other methods. DNS achieves competitive performance for all three dynamical processes and outperforms the STM-based ERTSS on some SMNR levels. A significant improvement can be seen in posterior estimation, particularly in low SMNRs where other methods perform poorly. At $-10$ dB SMNR, the iDANSE and DKS systems exhibit ALP values several magnitudes worse than DNS; for example, for the Chen system, DKS and iDANSE achieve $-3.2\times 10^4$ and $-1.1\times 10^{26}$ respectively, whereas DNS achieves $-88.95$). For iDANSE, the poor posterior estimation performance could be explained by overconfidence in state estimation. A visualization of an estimated state sequence for a stochastic Lorenz system at 0 dB SMNR is shown in Fig~\ref{fig:DNS_demonstration_for_Lorenz}. Then Fig~\ref{fig:uncertainty_comparison} provides a visualization of the trajectories of the first coordinate element of the three-dimensional state of Lorenz system.

We finally perform an ablation study of the skip connection in DNS. The results for the Lorenz system can be seen in Table~\ref{tab:ablation}. NMSE performance is lower at $10$ dB SMNR for the DNS without the skip connection. However, at low SMNR conditions, the opposite can be seen. Thus, having a skip connection of the states might lead to less overconfidence in poorly estimated states, improving performance at low SMNR levels.

\begin{table}
\centering
\renewcommand{\arraystretch}{0.8}
\caption{Ablation study of skip connection in DNS}
\small
\setlength{\tabcolsep}{1pt}
\begin{tabular}{l|ccc}
\toprule
 & \multicolumn{3}{c}{\textbf{Lorenz Attractor}} \\
 & \multicolumn{3}{c}{\footnotesize{SMNR (dB)}} \\
\textbf{Model} & -10 & 0 & 10 \\
\midrule
DNS \, & $\mathbf{-6.26}{\scriptstyle\pm 0.69}$ & $\mathbf{-16.84}{\scriptstyle\pm 0.31}$ & $-22.74{\scriptstyle\pm 0.53}$ \\[0.5ex]
Without Skip  & $-2.98{\scriptstyle\pm 1.24}$ & $-14.56{\scriptstyle\pm 1.71}$ & $\mathbf{-23.91}{\scriptstyle\pm 0.30}$ \\
\bottomrule
\end{tabular}
\label{tab:ablation}
\end{table}

\section{Conclusion}
In this article, we developed data-driven nonlinear smoother (DNS) for a model-free process. DNS is shown to be most effective in low signal-to-measurement-noise conditions, achieving orders of magnitude better posterior estimation than existing methods such as iDANSE and DKS. The use of skip connections is found to be effective in developing the deep recurrent architecture. Future works include extensions to real-world problems.

\vfill\pagebreak

\section{Acknowledgement}
The research is supported by funding from \href{https://www.digitalfutures.kth.se/}{Digital Futures Center}, \href{https://defence-industry-space.ec.europa.eu/system/files/2023-06/REACTII-Factsheet_EDF22.pdf}{European Defence Fund REACT II} project, and partially supported by the Wallenberg AI, Autonomous Systems and
Software Program (WASP) funded by the Knut and Alice Wallenberg Foundation. The computations were enabled by resources provided by Chalmers e-Commons at Chalmers.

\bibliographystyle{IEEEbib}
\bibliography{strings,refs}

\end{document}